\documentclass{elsart}
\usepackage{epsfig}
\usepackage{rotate}
\usepackage{amssymb}
\usepackage{amsmath}

\newcommand\pp{{\gamma '}}
\newcommand\g{{\gamma }}
\newcommand\gpp{{\gamma \to \pp }}
\newcommand\ppg{{\pp \to \gamma }}

\begin{document}


\begin{frontmatter}
\begin{center}
{\large \bf On search for hidden sector photons in  Super-Kamiokande }
\end{center}
\vspace{0.5cm}

\begin{center}
S.N.~Gninenko\footnote{E-mail address:
 Sergei.Gninenko\char 64 cern.ch}\\
{\it Institute for Nuclear Research of the Russian Academy of Sciences,\\
Moscow 117312}
\end{center}

\begin{abstract}
If hidden sector photons($\pp$)  exist, they 
could be produced through oscillations of  photons emitted by
 the Sun. We show that a search for these particles could be 
performed in Super-Kamiokande due to the presence in this detector 
of a large number of 
photomultiplier's (PMTs) with a relatively low noise and big size.    
The $\pp$s would penetrate the Earth shielding and would be 
detected by PMTs
 through their oscillations into real photons inside the PMTs vacuum volume. 
This would results in an increase of the 
PMT counting rate and it daily variations depending on the Earth position with respect to the Sun. The proposed search for this effect is sensitive to 
the $\gamma - \pp$ mixing  strength  as small as  
$\chi \lesssim 10^{-6}$ for the  $\pp$ mass region 
$10^{-3} \lesssim m_\pp \lesssim 10^{-1}$ eV and, in the case of 
nonobservation, could improve limits recently 
obtained from  photon regeneration laser experiments  for this 
mass region.
\end{abstract}
\end{frontmatter}

Several interesting extensions of the Standard Model (SM)  suggest
the existence of ''hidden'' sectors consisting of $SU(3) \times SU(2) \times U(1)$
singlet fields. These  sectors   of
particles do not interact with the ordinary matter directly and  couple to 
it by gravity and possibly by other 
very week forces.  If the mass scale of a hidden sector is too high, it is 
experimentally unobservable and indeed is hidden. However, there is  
 a class of models  with 
at least one additional  U'(1) gauge factor where the corresponding hidden 
gauge boson could be light. For example,  Okun  \cite{okun} proposed 
a paraphoton model with a massive hidden photon  mixing 
with the ordinary photon resulting in various interesting phenomena.
A similar model  of photon oscillations has been considered by 
Georgi et al. \cite{georgi}.   
Holdom \cite{holdom} showed, that  by adding a second,
massless  photon one could construct
grand unified models which contain particles with an electric
charge very small compared to the electron charge \cite{holdom}.
These considerations have stimulated new theoretical works
and experimental tests  reported in \cite{foot1}-\cite{moh1}
(see also  references therein).

In the Lagrangian describing  photon- hidden photon  system the so-called
 kinetic  mixing term  is given by \cite{okun,holdom,foot1} 
\begin{equation}
 L_{int}= -\frac{1}{2}\chi F_{\mu\nu}B^{\mu\nu} 
\label{mixing}
\end{equation}
where  $F^\mu$, $B^\mu$ are the ordinary 
 and the  hidden photon  field strength, respectively.
In the case when U'(1) is a broken symmetry, this kinetic mixing
can be diagonalized resulting in a non-diagonal mass term that 
mixes photons with hidden-sector photons. Hence, photons may oscillate 
into hidden photons, similar to vacuum neutrino oscillations.
Note that in the new field basis the ordinary photon remain unaffected, 
while  the hidden-sector photon ($\pp$) is completely decoupled, i.e. do not interact with the ordinary matter at all \cite{okun,holdom,foot2}.

Experimental bounds on $\pp$s
 can be obtained  from searches for the electromagnetic fifth force, 
\cite{okun,c1,c2}, from   
stellar cooling considerations \cite{seva1,seva2}, and from 
experiments using the method of  photon 
regeneration \cite{phreg}-\cite{pvlas}.
Recently, 
 new constrains on the $\g - \pp$ mixing strength  for the $\pp$ mass region 
$10^{-4}<m_\pp<10^{-2}$ have been
 obtained \cite{ring7} from the results of laser experiments
BMV \cite{bmv} and  GammeV \cite{gv}, looking for the photons regeneration 
through their interactions with light axion-like particles. The 
new results are a factor two better then the results obtained from the 
BFRT experiment \cite{bfrt}.    
The Sun energy loss arguments has also 
been  recently reconsidered \cite{red}. It has been pointed out
 that helioscopes searching for solar axions are sensitive to the solar
flux of hidden photons and the new limits on the $\g - \pp$ mixing parameter 
for the mass range  $m_\pp \simeq 0.01 - 1 $ eV obtained from the 
recent results of the  CAST collaboration
 \cite{cast1,cast2} have been reported \cite{red}. 
Strong bounds on new physics with $\pp$ at low energy scale 
 could be obtained  from astrophysical 
considerations \cite{blin}-\cite{dub}.
  However, such astrophysical constraints can be relaxed or evaded in 
some models, see e.g. \cite{masso}.
Hence, it is important to perform independent  tests on the 
existence of such particles in new experiments such, for example,  as 
ALPS \cite{alps}, LIPSS \cite{lipss}, OSQAR \cite{osqar}, and PVLAS LSW 
\cite{lsw}.\\ 
In this note,  we propose  a direct experimental search for $\pp$ particles
 which might be present in the photon flux from the Sun.
The experiment could be performed with  the Super-Kamiokande neutrino 
detector and is based on the  photon-regeneration method, used at low energies.


If $\pp$s can be produced through oscillations of real photons, 
it is naturally 
to consider the Sun as a source of low energy $\pp$s. Indeed,   
 it is well known that the total emission rate of the Sun 
  is of the order of $3.8\times 10^{26}$ W. 
 The emission photon energy spectrum is well understood. It has a
broad distribution over energies up to
10 keV,  and corresponds 
roughly to the black-body radiation at the temperature $\simeq 5800$ K.
The maximum in the solar emission spectrum is at about 500 nm,
 in the blue-green part of the visible spectrum. 
Direct visible photons arrive at the Earth at a rate of $\simeq 4\times10^{17}/cm^2/s$. 
Thus, if $\gpp$ transitions occur it might be  advantageous to
search for hidden sector photons through regeneration of real photons  
using the visible part of the emission spectrum from the Sun, where the 
photon flux is maximal.

Among detectors suitable for such kind of search the most promising one is  
Super-Kamiokande (SK) \cite{sk}. This is a large, underground, water Cherenkov detector located in a mine in the Japanese Alps \cite{sk}.  
The inner SK detector is a  tank, 40 m tall by 40 m in diameter.
 It is filled with 5$\times 10^4$ m$^3$ of ultra-pure water, 
the optical attenuation length $L_{abs}\gtrsim$ 70 m, and 
is viewed by 11146 photomultiplier tubes (PMT) with 
7650 PMTs mounted on a barrel (side walls) and 3496 PMTs on the top
and bottom  endcaps.

\begin{figure}
\begin{center}
  \epsfig{file=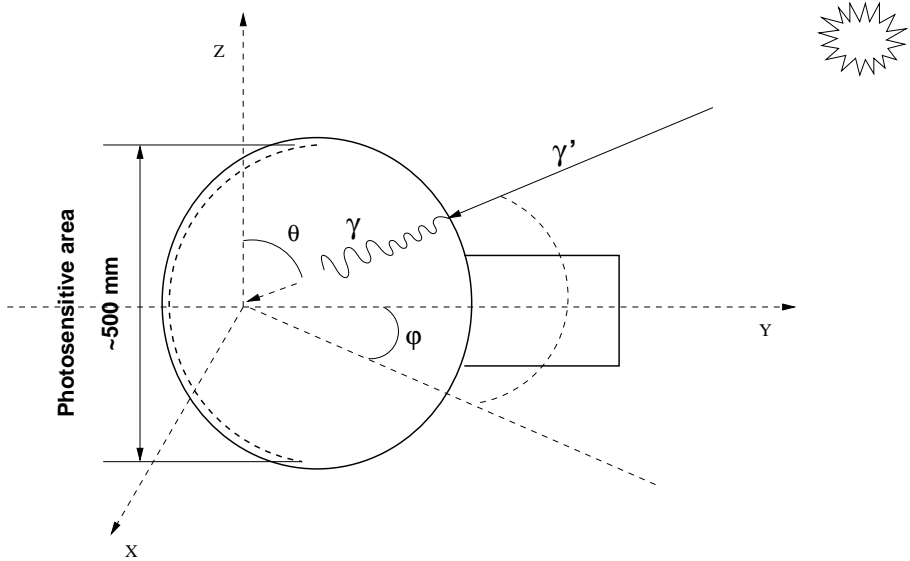,width=120mm}
\end{center}
 \caption{\em Schematic illustration of the direct search for light 
hidden-sector photons in the Super-K experiment.
Hidden photons penetrate the Earth and convert into visible photons
inside the vacuum volume of the Super-K PMTs. 
This results in an increase of 
the counting rate of those Super-K PMTs that are 
''illuminated'' by the Sun from the back, 
in comparison with those facing the Sun. If,  
for example, the Earth rotates around the Z-axis, 
the counting rate is  a periodic function of the angle $\psi$, i.e. is daily  
modulated. } 
\label{superk_PM}
\end{figure}
The PMTs (HAMAMATSU R3600-2) have  $\simeq$ 50 cm 
in diameter\cite{sk}. The full effective PMT photocathode coverage of the 
inner detector surface is 40\%.
The  photo cathode, the dynode system and 
the anode are located inside a glass envelope serves as a pressure boundary 
to sustain high vacuum conditions inside the almost spherical shape PMT.  
The photocathode is made of bialkali (Sb-K-Cs) that 
matches the wave length of Cherenkov light. The quantum efficiency 
is $\simeq$ 22\% at the 
typical wave length of Cherenkov light $\simeq$ 390 nm. 
For the search for $\g - \pp$ oscillations it is important
to have the ability to see a single photoelectron (p.e.) peak,
 because the number of photons arriving at a PMT is exactly one.  
 The single p.e. peak is indeed clearly seen (see e.g., Figure 9 in 
Ref. \cite{sk}) allowing to operate PMTs
in the SK experiment at a low threshold equivalent to 0.25 p.e.. It is also 
important, that 
the average PMT dark noise rate  at this threshold is just about 3 kHz.

If $\pp$s are long-lived noninteracting particle, they  would   
 penetrate the Earth shielding   
and oscillate into real photons in the free space 
between the PMT envelope  and the  photocathode, as shown in 
Figure \ref{superk_PM}. The photon then would convert in the
photocathode into a single 
photoelectron which would be detected by the PMT.  
Thus, the effect of $\pp \rightarrow \gamma$ oscillations
could be searched for in the SK experiment through 
 an increase of the counting rate of 
those PMTs that are ''illuminated '' by the Sun from the back, as shown in 
Figure \ref{superk_PM},  
in comparison with 
those facing the Sun. The increase of the counting rate in a particular
PMT depends on its orientation with respect to the Sun and is daily 
modulated. Therefore, the overall counting rate of events  from  
$\pp \rightarrow \gamma$ oscillations could also be daily modulated 
depending on the local 
SK position with respect to the Sun and  the Earth rotation axis.

The number $\Delta n_\g$ of expected signal events from $\ppg$ conversion  in SK  is given by the 
 following expression:   
\begin{equation}
\Delta n_\g = \Sigma_1^N \int_{\omega_1}^{\omega_2} I_\g(\omega) \cdot \eta(\omega) \cdot P_{\g \to \pp}(m_\pp, \omega) \cdot P_{\pp \rightarrow \g}
(m_\pp,\omega)\cdot  d \omega d\Omega 
\label{s1}
\end{equation}
Here, N is the number of SK PMTs, $I_\g$ and $\omega$ are the photon flux and  energy, respectively, $\eta$
is the detection efficiency, and 
$P_{\g \to \pp}(m_\pp,\omega)$ and $P_{\pp \rightarrow \g}(m_\pp, \omega)$
are  the $\g \to \pp$ and $\pp \to \g$ transition probabilities, respectively,
given by: 
\begin{equation}
P_{\g \to \pp}(\omega, m_{\pp}) = 4\chi^2 sin^2\Bigl(\frac{\Delta q L}{2}\Bigr) 
\label{prob1}
\end{equation}
\begin{equation}
P_{\pp \to \g}(\omega, m_{\pp}) = 4\chi^2 sin^2\Bigl(\frac{\Delta q l}{2}\Bigr)  
\label{prob2}
\end{equation}
where $L, l$ are the distances between the Sun and the Earth,  and
between the $\pp$ entry point to the PMT and the PMT photocathode, 
respectively, and $\Delta q$ is
the momentum  difference between the photon and hidden photon:
\begin{equation}
\Delta q = \omega - \sqrt{\omega^2 - m_{\gamma '}^2} \approx \frac{m_{\gamma '}^2}{2\omega}
\end{equation}
assuming $m_{\gamma '}\ll \omega$. 
In the absence of photon absorption,
the  maximum of the $\gamma \to \pp (\ppg)$ transition probability
at a distance $l$  corresponds to the case when $|\Delta q l| \simeq \pi$. 
When $|\Delta q l| \ll \pi$ the photon and the hidden photon fields 
remain in phase and propagate coherently over the length $l$. In this 
case the transition probability is proportional to $l^2$.
For $\omega \simeq 3 eV$ and for the maximum distance $l\simeq 50$ cm, 
the smallest $\pp$-mass which can be effectively  
explored at the Super-K detector with the 
photon-regeneration method is  $m_\pp \simeq 10^{-3}$ eV. For the mass range 
$m_\pp \gtrsim 10^{-3}$ eV and for $L\simeq 10^{13}$ cm,  
\begin{equation}
\frac{m_\pp^2 L}{4 \omega} \gg 1
\end{equation}
thus, the sinus of Eq.(\ref{prob1}) is averaged to 1/2 and Eq.(\ref{s1}) 
reads
\begin{equation}
\Delta n_\g =8 \chi^4 \Sigma_1^N \int_{\omega_1}^{\omega_2} I_\g(\omega) \cdot \eta(\omega) \cdot  sin^2\Bigl(\frac{m_\pp^2 l}{4\omega}\Bigr) d \omega d\Omega 
\label{s2}
\end{equation}

The significance $S$ of the $\pp$  discovery  with the Super-K detector 
scales as  \cite{bk}
\begin{equation}
S=2(\sqrt{\Delta n_\g + n_b}-\sqrt{n_b}) 
\end{equation}
where $n_b$ is the number of expected background events.
The excess of $\ppg$ events in the Super-K detector can be 
calculated from the result of a numerical integration of Eq.(\ref{s2})
over photon trajectories pointing to the PMT. 
In this calculation we use a simple model of  PMTs, 
 shown in Figure \ref{superk_PM}, 
without taking into 
account the PMT internal structure and  dead materials which might results
 in some reduction of the signal due to the photon absorption and 
damping of $\pp -\g$ oscillations.  
The Sun emission spectrum is described by the Planck's low 
of black-body radiation.   
We also assume that the Sun is located in the plane $\Theta = \pi/2$ and 
the Earth rotates around the $Z$-axis, which is the local vertical in SK, 
  see Figure \ref{superk_PM}.
For the considered  mass  range  $10^{-3}<m_\pp<10^{-1}$
the shape of the $\pp$ energy spectrum arriving at the Earth
is the same as for solar photons
 due to the large distance between the Sun and the Earth.
In the PMT vacuum volume
 not all $\pp$ energies effectively contribute 
to the signal because of its sinus dependence on $\Delta q$ and $l$, see 
Eq.(\ref{prob2}). The total number of expected events from $\g - \pp$ 
oscillations collected  in the SK experiment during the  exposition time
 $t$ is estimated for the solar flux $I_\g \simeq 1.4 \times 10^3$ W$/m^2$ and
 taking into account  the overall detection efficiency, which includes 
the quantum efficiency of the photocathode, the
secondary electron collection efficiency, the detection threshold, and the
geometrical acceptance. 
\begin{figure}[htb!]
\begin{center}
  \epsfig{file=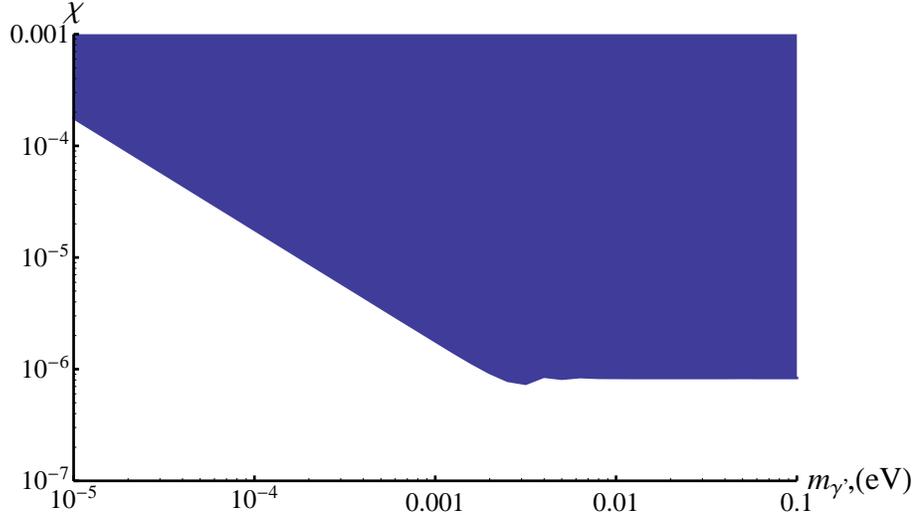,width=120mm}
\end{center}
 \caption{\em The region in the ($m_\pp, \chi$) plane (filled area)
 which could be excluded by the proposed experiment.}
\label{plot}
\end{figure}
 Assuming 
the main background source is the PMT dark noise gives
\begin{equation}   
 n_b = n_0 N' t
\end{equation}
Here $N'$ is the number of SK  PMTs contributing to the signal, and
$n_0 \simeq 3$ kHz is the average background counting rate of the PMTs 
\cite{sk}. Finally, 
taking $S=3$, $ N'\simeq 7 \times 10^3$, and  $ t\simeq 10^7$ s results in
 the exclusion region in the  ($m_\pp, \chi$) 
plane  shown in Figure \ref{plot}.
For the mass region $m_\pp \gtrsim 10^{-3}$ eV  the  limit for the 
mixing parameter is 
\begin{equation}
\chi< 8.3 \times10^{-7} 
\label{limit}
\end{equation}

The limit estimated for the
 configuration when the Sun is located at $\Theta = 0$ is comparable 
than the one of Eq.(\ref{limit}).\\
We see that the sensitive search of $\pp$'s in the SK experiment is   
possible due to unique combination of several factors, namely, 
i) the presence of the large number of PMTs with
a relatively large free vacuum volume; ii) the high efficiency of the single 
photon detection; and iii) the relatively low  PMT dark noise.
 The statistical limit on the sensitivity of the proposed experiment 
is set by the number of PMTs and by
the value of the dark noise ($n_0$) in the SK detector. 
The systematic errors the are not 
included in the above  estimate, however they could be reduced by the precise 
monitoring of the PMTs gain \cite{sk}. For the mass region 
$10^{-3} \lesssim m_\pp \lesssim 10^{-1}$ eV the estimated upper limit of Eq.(\ref{limit}) is slightly better then the
 limits recently obtained by  Ahlers et el. \cite{ring7} from 
 laser experiments. 
This estimate may be strengthened  by more accurate and  detailed Monte Carlo
simulations of the proposed experiment.

Finally note, that in the presence of photon absorption, 
the probability for a hidden  photon 
to pass through a conversion region of a length $L'$ and to arrive to a 
detector as an ordinary photon is given by \cite{red}, see also
\cite{rs,vanb}:
\begin{equation}
P_{\pp \to \g}(\omega, m_{\pp}) = \frac{\chi^2 m_\pp^4}{(m_\g^2 - m_\pp^2)^2+(\omega \Gamma)^2} \Bigl(1+e^{-\Gamma L} - 2 e^{- \Gamma L'/2} cos \Delta q L'\Bigr)
\label{abs}
\end{equation}

In the above formula,  $\Delta q=(m_\gamma^2 - m_\pp^2)/2\omega$ is 
the difference between  the $\pp$ and the  photon momenta in the medium
for a photon energy $\omega$;
$m_\gamma$ plays the role of the plasma frequency (photon mass) in a target medium:
$m_\gamma^2 = 4\pi\alpha N_e / m_e$ with
 $N_e$  the electron number density and $m_e$ the electron mass. 
For the SK water target  
$m_{\gamma}\simeq 20$ eV, and the  photon absorption 
rate  is $\Gamma \simeq 1/L_{abs} \simeq 3\times 10^{-9}$ eV. 
Thus, although  photon absorption in the target is small, 
 the transition probability of Eq.(\ref{abs}) is suppressed 
for the mass region $m_\pp \ll 1$ eV by the high photon 
mass in the target medium. Even in the  case when the SK tank is 
filled with air,  the target photon mass is $m_\g \gtrsim 10^{-2}$ eV  and the sensitivity 
of the experiment cannot  be improved and extended to the smaller $\pp$ 
mass region  by searching $\g - \pp$ oscillations in the inner SK detector
volume.

{\large \bf Acknowledgments}

The author thank  N.V. Krasnikov and V. Popov for useful
discussions. Help of A. Korneev in calculations is
 gratefully acknowledged.

\end{document}